\documentstyle[a4,12pt]{article}

\newcommand{\ir}{infra-red}

\newcommand{\lgl}{light gluino}

\newcommand{\sm}{standard model}
\newcommand{\mssm}{minimal supersymmetric \sm}
\newcommand{\lsp}{lightest supersymmetric particle}
\newcommand{\xs}{cross section}
\newcommand{\br}{branching ratio}

\newcommand{\EW}{electroweak}

\newcommand{\susy}{supersymmetry}

\newcommand{\trm}{transverse momentum}

\newcommand{\co}{chargino}

\newcommand{\no}{neutralino}

\newcommand{\ep}{\mbox{$e^+e^-$}}

\def\lr3{$SU(3)_L\otimes SU(3)_R$}

\def\z0{$Z^0$}
\def\Z0{$Z^0$}

\def\ep{$e^+e^-$}

\def\gsim{\buildrel{\lower.7ex\hbox{$>$}}\over{\lower.7ex\hbox{$\sim$}}}
\def\lsim{\buildrel{\lower.7ex\hbox{$<$}}\over{\lower.7ex\hbox{$\sim$}}}

\newenvironment{comment}[1]{}{}

\newcommand{\beq}{\begin{eqnarray}}
\newcommand{\eeq}{\end{eqnarray}}

\begin{document}

\thispagestyle{empty}
\setcounter{page}{0}

\begin{flushright}
MPI-Ph/93-69\\
LMU-15/93\\
September 1993
\end{flushright}
\vspace*{\fill}

\begin{center}
{\Large\bf Light Gluinos at LEP}\\
\vspace{2em}
\large
\begin{tabular}[t]{c}
Frank Cuypers\\
\\
{\it Max-Planck-Institut f\"ur Physik, Werner-Heisenberg-Institut,}\\
{\it D--80805 M\"unchen, FRG}\\
\end{tabular}
\end{center}
\vspace*{\fill}

\begin{abstract}
\noindent
We study the production and decay of \lgl s
in \ep\ collisions.
We suggest a signature
which suffers little from backgrounds
and argue that the \lgl\ window
can already be closed with the existing LEP data,
provided the gluino lifetime is such that
it decays within the detectors.
\end{abstract}
\vspace*{\fill}

\newpage

\section{Introduction}

Historically seen,
there are three kinds of elementary particles:
those whose discovery was unexpected
and revolutioned our perception of the world,
like the neutron or the strange quark;
those whose existence had been foreseen
and whose discovery confirmed the speculations of a brilliant mind,
like the positron or the \EW\ gauge bosons;
but there are also those which were predicted
and have never been found.
Very often it is difficult to preclude the existence of the latter
on the sole basis that they have not shown up yet in experiments.
Indeed,
generally the models which predict their existence
can be adapted in such a way
that their couplings to ordinary matter
becomes so low
that even the highest precision measurements
would be unable to detect their presence,
or that the mass of these particles
becomes so high
that even the highest energy experiments would not produce them.
This adaptation of the models often goes with a dramatic decrease
in the esthetic value of the considerations
which,
no doubt,
motivated their elaboration.
In some cases,
a particle ``soon-to-be-discovered''
is even demoted to an ``invisible'' status.

There is,
however,
an exception to this rule:
the \lgl.
The gluino is predicted by the theory of \susy\
to be the spin 1/2 partner of the gluon.
Like the gluon,
it is its own anti-particle,
hence a Majorana spinor,
and it couples to ordinary matter with the same strength as a gluon.
Strangely enough,
in spite of its strong interactions
and substantial efforts from both theorists and experimentalists,
it has not yet been possible to rule out
the existence of a gluino
weighing less than 5 GeV \cite{pdg,LG2,LG3,LG6,LG1}!
A number of studies has been devoted
to exclude at least some windows
within this impertinent mass gap,
but none of them has been absolutely conclusive.

It is the purpose of this letter
to attempt to settle the ongoing controversy
about the existence of this \lgl.
For this
we compute the production rate of gluinos
in \ep\ collisions
and study their decay signature.
It turns out that the background
arizing from the \sm\ and detector inefficiencies
can be virtually eliminated with $b$-tagging.
We confined this study to the case of LEP,
because of the high \xs s obtained on the $Z^0$ peak.

\section{Gluino Production}

The lowest order \xs\ for producing gluinos in \ep\ collisions
can be computed from the Feynman diagrams of Fig.~\ref{feynprod}.
For massless quarks
it can be written
\beq
\sigma_{\tilde g}
&=&
{2N_c\alpha^2\alpha_s^2 \over 3\pi s}
{}~
\sum_f\Gamma_f
{}~
\int^s_{4m^2_{\tilde g}}
dm^2
{}~
\sqrt{1-{4m^2_{\tilde g}\over m^2}}
\left( 1+{2m^2_{\tilde g}\over m^2} \right)
{1 \over m^2}
{}~
T\left( {m^2\over s} \right)
\label{tot20}\ ,
\eeq
where $N_c=3$ is the number of colours,
$\alpha$ is the fine structure constant,
$\alpha_s$ is the strong coupling constant
and $m_{\tilde g}$ is the gluino mass.
The scaled $\gamma$--$Z^0$ propagator squared $\Gamma_f$
is summed over all active quark flavours $f$.
It is given by
\beq
\Gamma_f
&=&
Q_f^2
-
v_e v_f Q_f {s(s-m_Z^2) \over (s-m_Z^2)^2 + m_Z^2\Gamma_Z^2}
+
(v_e^2+a_e^2) (v_f^2+a_f^2) {s^2 \over (s-m_Z^2)^2 + m_Z^2\Gamma_Z^2}
\label{tot40}\ ,
\eeq
where $v_e,~a_e,~v_f$ and $a_f$
are the electron's and quark's vector and axial vector couplings
to the $Z^0$ boson
and $Q_f$ is the quark's charge.
The integral in Eq.~(\ref{tot20}) is over the gluon virtuality
and the function $T$ is given by
\beq
T(x)
&=&
{2\over3}
\bigg\{
	(1+x)^2
	\left[
		\ln^2 \left(x+1/x+2\right) - \pi^2/3 + \Sigma(x)
	\right]
\nonumber\\&&\quad
	+ \left( 3+4x+3x^2 \right) \ln x + 5 \left(1-x^2 \right)
\bigg\}
\label{tot50}\ ,
\eeq
where
\beq
\left\{
\begin{array}{l}
	\Sigma(x) = \displaystyle\sum^\infty_{n=1} ~ c_n
	~ \left( \displaystyle{4\over x+{1/x}+2} \right)^n
	\\
	c_n = c_{n-1} ~ \displaystyle{2n-1\over2n-2} \left( {n-1\over n} \right)^3
	\\
	c_1 = 1
\end{array}
\right.
&\Rightarrow&
\left\{
\begin{array}{l}
	\Sigma(0) = 0
	\\
	\Sigma(1) = \pi^2/3 - \ln^2 4
\end{array}
\right.
\nonumber
\eeq

For low gluino masses
the gluon virtuality $m^2$ in Eq.~(\ref{tot20})
is allowed to approach zero.
In this limit
the function $T$ takes the asymptotic divergent form \cite{SW}
\beq
T(\epsilon)
=
{2\over3} \ln^2 \epsilon + 2 \ln \epsilon + {10\over3} - {2\over9}\pi^2
&\qquad {\rm for} \qquad&
\epsilon=0
\label{tot60}\ .
\eeq
This expression clearly displays
the double and single logarithmic divergences
which appear when integrating over the \ir\ and collinear
regions of phase space.
These are approached
only when the gluon virtuality in Eq.~(\ref{tot20}) can be small,
{\em i.e.} when the gluino mass is small.
\begin{comment}{
This effect
is clearly shown by the behaviour of the upper curve of Fig.~\ref{masstot}
at low values of $m_{\tilde g}$.
}\end{comment}
The \xs\ obtained from Eq.~(\ref{tot20}) is thus not trustworthy
when the gluino is too light.
A convenient method to counter this deficiency
is to avoid alltogether
the \ir\ and collinear regions of phase space
where the \xs\ is unphysically large.
The easiest kinematical cut which comes to mind
is to impose a lower bound $M$
on the invariant mass of the gluino pair.
This is done in Eq.~(\ref{tot20})
by replacing the lower integration limit
$4m^2_{\tilde g}$ by $M^2$.
This way,
the virtuality of the gluon is never allowed to come close to zero
and the doubtful regions of phase space are not approached.
\begin{comment}{
The resulting \xs\
is shown in Fig.~\ref{masstot}
as a function of the gluino mass
for several choices of the minimum invariant mass $M$ of the gluino pair.
}\end{comment}

Of course,
one might wonder how to implement this cut experimentally.
Indeed,
the produced gluinos,
and partons in general,
radiate soft and collinear gluons
which themselves radiate even softer gluons.
This cascade develops until the relative \trm\ of the partons
reaches the hadronization scale ($\sim1$ GeV)
and a jet is born.
What happens then
can only be simulated with models
({\em e.g.} string or cluster hadronization \cite{hadronization}).
Something seems to be garanteed,
however:
heavy quark jets contain a heavy hadron,
light quark and gluon jets contain only light hadrons,
and gluino jets contain a glueballino.
This makes $b$-tagging an invaluable tool for recognizing
quark jets as such.
With this option
the kinematical cut can be unambiguously implemented
on the overall invariant mass
of the remaining jet or jets,
those which might have been initiated by a pair of gluinos.

Note that Eq.~(\ref{tot20}) also gives the rate at which
two quark pairs $q\bar qq'\bar q'$ are produced,
if the the colour factor $N_c$ is replaced
by the number of active flavours $N_f$.

\section{Gluino Decay}

We shall assume in the following that
R-parity remains unbroken
and that the \lsp\ is a \no.
If this is the case
a \lgl\ decays predominantly
via the exchange of a squark
into a quark-antiquark pair
accompanied by the lightest \no,
which remains undetected.
\begin{comment}{
The lowest order Feynman diagrams
depicting this process
are shown in Fig.~\ref{feyndec}.
}\end{comment}
Higher order decay mechanism
can compete only for very contrived values
of the \susy\ parameters \cite{HKN}\
and we therefore do not consider them here.

According to this scenario,
if the gluino is very light,
the lightest \no\ has to be almost massless.
This can only be the case
if either of the \susy\ parameters $\mu$ or $M_2$ is small.
In turn,
$\tan\beta$ can then not be much larger than one,
in order to accomodate the LEP bounds on the \co\ mass.
If the light mass of the \no\ is achieved
by a low value of $\mu$,
its main component is a Higgsino.
In this case,
the gluino will eventually decay,
but its lifetime is too long for this to happen
within a detector.
On the other hand,
if $M_2$ is small,
the lightest \no\ is dominantly a photino
and the lifetime of the gluino
is of the order of a weak decay lifetime.
We therefore assume in the following that
\beq
M_2 = 0
& \qquad \Rightarrow \qquad &
\left\{
\begin{array}{l}
	\tilde\chi_1^0 = \tilde\gamma
	\\
	m_{\tilde\gamma} = 0
\end{array}
\right.
\label{life10}\ .
\eeq

For $N_f$ flavours of massless quarks and heavy squarks,
the decay width of a free gluino is \cite{LG4}
\beq
\Gamma
&=&
{\alpha\alpha_s\over48\pi}
\quad
\sum_f^{N_f} Q_f^2
\quad
{m_{\tilde g}^5\over m_{\tilde q}^4}
\label{life20}\ ,
\eeq
where $\alpha$ is the fine structure constant,
$\alpha_s$ is the strong coupling constant,
$Q_f$ is the quark charge
and $m_{\tilde g}$ and $m_{\tilde q}$ are the gluino and squark masses.
\begin{comment}{
The corresponding lifetime $\tau = \hbar/\Gamma$ of the gluino
is plotted in Fig.~\ref{lifetime}
as a function of the gluino
for several values of the squark mass.
The kink at 2.70 GeV
is due to the opening of the charm channel.
The dotted lines on Fig.~\ref{lifetime}
show the average traveling distance inside a detector
\beq
<L>
&=\quad
<\gamma\beta> c
\quad\displaystyle{\hbar\over\Gamma}
&=\quad
\sqrt{{E^2\over m_{\tilde g}^2} -1} \quad c \quad {\hbar\over\Gamma}
\label{life30}\ ,
\eeq
taking here $E = \sqrt{s}/4 = m_Z/4$.
}\end{comment}
The invariant mass $M_h$ of the quark pair
emerging from the decay of the gluino
is distributed according to
\beq
{1\over\Gamma}
{d\Gamma\over dM_h}
&=&
{4\over m_{\tilde g}^8}
M_h \left( m_{\tilde g}^6 - 3m_{\tilde g}^2M_h^4 + 2M_h^6 \right)
\label{life40}\ .
\eeq
\begin{comment}{
This distribution is shown in Fig.~\ref{hadmass}
for different choices for the gluino mass.
}\end{comment}

All this was said for a free gluino.
In reality
a gluino would emerge at the end of a hadronization process
in a colour blanched bound state,
most probably a glueballino $\tilde G = g\tilde g$.
\begin{comment}{
The decay mechanism of this gluon-gluino meson
is sketched in Fig.~\ref{hadronization}.
Its decay mechanism resembles very much a hadronic semi-leptonic decay,
except for the absence of a lepton.
}\end{comment}
The decay of this gluon-gluino meson
is dictated by the decay mechanism of the gluino.
It consists of a secondary vertex within a jet,
to which no charged track is leading.
Off this vertex,
only hadrons emerge
and their invariant mass is continuously distributed
over the full range $[0,m_{\tilde G}]$
according to Eq.~(\ref{life40}).
Actually,
bound state effects are expected to harden this free gluino spectrum.
However,
since this shift towards higher invariant masses
amounts to only a small correction in the present analysis,
we assume here the validity of Eq.~(\ref{life40})
(with $m_{\tilde G}$ replacing $m_{\tilde g}$)
also for the hadron spectrum of a glueballino.
\begin{comment}{
Within this approximation
Fig.~\ref{hadmass} also shows
the invariant mass distributions
of the hadrons in glueballino decays.
}\end{comment}
Only the long-lived neutral hadrons $(K^0,~\Lambda,~D^0,~B^0,\dots)$
have similar signatures,
but their decay hadrons emerge with sharply peaked invariant masses.

At this stage
the lifetime of a glueballino
can only be estimated.
According to Ref.~\cite{LG4}\
it is approximated by the same formula (\ref{life20})
as for the decay of a free gluino,
replacing the gluino mass
by an effective glueballino mass
$m^*_{\tilde G} \approx .75 m_{\tilde G}$.
\begin{comment}{
The curves in Fig.~\ref{lifetime} remain thus valid here,
provided the gluino masses on the x-axis
are correctly reinterpreted as $m^*_{\tilde G} = 3/4 m_{\tilde G}$.
}\end{comment}
But the glueballino mass itself
is also just a guess.
Nevertheless,
even for a massles gluino
it is unlikely to be smaller than 1 GeV,
and for a heavier gluino,
it will probably be close to the mass of the gluino itself.
For our purposes
we assume that the effective mass to be used in Eq.~(\ref{life20})
is well approximated by the mass of the gluino itself,
but cannot be less than .75 GeV.
The range of values which
the squark and effective glueballino masses can then take
for the glueballino decay to be detectable
within a typical time projection chamber ($L\lsim2$m)
or vertex detector ($L\gsim2$mm) 
is shown in Fig.~\ref{mgmsq}.
This fills a large portion of the parameter space left
by some other conclusive studies \cite{pdg}\
or applicability arguments:
$74~{\rm GeV} < m_{\tilde q} < 2~{\rm TeV}$
and
$.75~{\rm GeV} < m^*_{\tilde G} < 4~{\rm GeV}$.
The latter bounds should correspond closely to
$0~{\rm GeV} < m^*_{\tilde g} < 4~{\rm GeV}$.
The kink at $m^*_{\tilde G} = 2.7$ GeV
is due to the opening of the charmed decay channel.

\section{Signal and Backgrounds}

The procedure we propose
for discriminating a \lgl\ in \ep\ collisions
consists of the following selection criteria:
\begin{itemize}
\item	Events with three or more jets.
\item	Two of these jets contain clearly identified $b$ quarks.
\item	The overall invariant mass of the remaining jet(s)
	exceeds a certain lower bound $M$.
\item	These remaining jet(s) contain two secondary vertices
	with only hadrons emerging
	and which are not initiated by a charged track.
\item	The invariant masses of the hadrons leaving these secondary vertices
	do not overlap with the masses of the long-lived
	neutral hadrons\footnote{
		There are more long-lived baryons,
		but their occurence is so rare
		that we can safely ignore them here.
	}: $K^0,\ D^0,\ B^0$ and $\Lambda$.
\end{itemize}
In order to implement the first two requirements
one should,
in principle,
define what is a jet.
Typically,
this can be done by invoking a jet-finding procedure,
like the Durham \cite{durham}\ or the Jade \cite{jade}\ algorithm.
A more complete study
including a full detector simulation,
should indeed include these refinements.
Here,
however,
we do not incorporate them,
because they cannot be applied
to our nearly integrated \xs\ (\ref{tot20})
and do not modify our conclusions.

\begin{comment}{
The incidence of the last criterium
is shown in Fig.~\ref{hadmass},
where the regions within 50 MeV of the masses
of the $K^0,~\Lambda,~D^0$ and $B^0$
are displayed.
Although this resolution is a blatant insult
to any of the LEP detectors,
we keep it this high to stay on the conservative side.
}\end{comment}

In principle, there is no \sm\ background for events
satisfying the five requirements listed above.
In practice,
however,
there are cracks in the detectors,
through which a hadron can escape
and falsify the invariant mass measurement
of a long-lived hadron.
Moreover,
some of these hadrons can also decay semi-leptonically,
in which case the invariant mass measurement
can also be falsified
if an electron is mistaken for a pion.

Obviously,
these effects can only be accurately estimated
with a complete and dedicated detector simulation.
Still,
an order of magnitude calculation
reveals that these backgrounds are negligible.
Indeed,
if the selection criteria above are implemented,
the heavy hadrons which are at the origin of the background
can only be produced in 4-quark events\footnote{
	The production of heavy flavours
	is very much suppressed in gluon jets.
	Therefore $b\bar bg$ or $b\bar bgg$ events cannot
	satisfy the selection criteria.
}
of the type $b\bar bq\bar q$.
The rate of these events is ${N_f \over N_c} \sigma_{\tilde g}$
(see Eq.~(\ref{tot20}),
where the only active flavour is $f=b$)
and the background \xs\ is thus
\beq
\sigma_{\rm SM}
\quad=\quad
\sigma_{\tilde g}~
{N_f\over N_c}~
\eta^2
& \Big[ &
	P(\Lambda)~~~ BR(\Lambda~ \to e^\pm\nu + {\rm hadrons})
\nonumber\\
& + &	P(D^0)~ BR(D^0 \to e^\pm\nu + {\rm hadrons})
\nonumber\\
& + &	P(B^0)~ BR(B^0 \to e^\pm\nu + {\rm hadrons})
\quad \Big]^2
\label{anal10} ,
\eeq
where $\eta$ is the probability to misidentify electrons.
The \br s $BR$ for electronic decays of the $\Lambda$, $D^0$ and $B^0$
are respectively .083\%, 7.7\%\ and 12.1\%\ \cite{pdg}
while the probabilities $P$ of finding these hadrons in a quark jet
are of the order of 30\%\ or less.
The $K^0$ plays no role here,
because the $K^0_L$ decays far outside the detectors
and the $K^0_S$ has no semi-leptonic decay.
Typically,
LEP detectors can discriminate electrons and pions
better than to 1\%\ $(\eta=10^{-2})$.
The ratio
$\sigma_{\tilde g} / \sigma_{\rm SM}$
is thus so large
that we can safely ignore the backgrounds
caused by inefficiencies in the electron identification.
The same argument holds for the backgrounds
due to imperfect detector hermeticities.

The gluino \xs\ which is obtained
when implementing our selection criteria
is displayed in Fig.~\ref{masscut}
as a function of the gluino mass
for several choices of the minimum invariant mass $M$ of the gluino pair.
To obtain this result,
we assumed an invariant mass resolution of 100 MeV
and chose the $b$-tagging procedure
to be 50\%\ efficient.
For 100 pb$^{-1}$ of integrated luminosity
(which should have been accumulated at LEP1 with $b$-tagging by now),
even a 20 GeV gluino
(which has already been clearly ruled out by previous experiments)
would provide more than 10 events
which cannot be explained within the framework of the \sm!
In turn,
a \lgl\ of less than 5 GeV
would generate several thousands of such events.

\section{Conclusions}

We have studied the production and decay of \lgl s at LEP
and suggested a signature with no or negligible backgrounds
from the \sm\ and detector inefficiencies.
It appears that the existence of a \lgl\
can easily be confirmed or ruled out
with the data already accumulated at LEP,
provided the glueballino $g\tilde g$ bound state
decays neither outside the detectors
nor too close to the primary vertex.
The corresponding domain of observability
in the space of the glueballino and squark masses
is shown in Fig.~\ref{mgmsq},
assuming the photino achieves its ``\lsp'' status
with $M_2\approx0$
within the framework of the \mssm.

\bigskip
\bigskip
I am very much indebted to
Louis Clavelli,
Alexander Khodjamirian,
Wolfgang Ochs,
Geert Jan van Oldenborgh,
Reinhold R\"uckl
and Ron Settles
for the precious discussions they have granted me.

\newpage

\begin{figure}[t]
\begin{center}
\begin{picture}(300,150)(0,0)
{\tt the .ps file of this preprint and its figures,
can be obtained via anonymous ftp
from 129.187.198.1 in /preprints/mpi9369.ps .}
\end{picture}
\end{center}
\caption{Lowest order Feynman diagrams contributing to gluino production
	in $e^+e^-$ collisions.}
\label{feynprod}
\end{figure}

\begin{figure}[t]
\centerline{
\begin{picture}(600,504)(0,0)
{\tt the .ps file of this preprint and its figures,
can be obtained via anonymous ftp
from 129.187.198.1 in /preprints/mpi9369.ps .}
\end{picture}
}
\caption{Curves corresponding to the average distances of 2 mm and 2 m
	traveled by a glueballino
	in the still allowed space of the glueballino effective mass
	and the squark mass.
	The darkened area can be explored by the method advocated here.}
\label{mgmsq}
\end{figure}

\begin{figure}[t]
\centerline{
\begin{picture}(1000,504)(0,0)
{\tt the .ps file of this preprint and its figures,
can be obtained via anonymous ftp
from 129.187.198.1 in /preprints/mpi9369.ps .}
\end{picture}
}
\caption{LEP cross sections for the gluino signal described in the text,
	as a function of the gluino mass.
	The incidence of different choices for the lower bound
	imposed on the invariant mass of the non-$b$ quark jets
	is also shown.}
\label{masscut}
\end{figure}

\end{document}